# Plataforma para visualização geo-temporal de apinhamento turístico

**SIMÕES, Rodrigo[1]; BRITO E ABREU, Fernando[2]; LOPES, Adriano[3]**

ISTAR-IUL; Instituto Universitário de Lisboa; Av. das Forças Armadas, 40, 1649-026, Lisboa
[1]rjbss@iscte-iul.pt; [2]fba@iscte-iul.pt; [3]adriano.lopes@iscte-iul.pt

**Resumo:** O apinhamento turístico degrada a experiência dos visitantes e impacta negativamente o ambiente e a população local, podendo tornar insustentável o turismo em destinos populares. Isto motivou-nos a desenvolver, no âmbito do projeto europeu *RESETTING* relacionado com a transformação digital do turismo, uma plataforma para visualizar este apinhamento, explorando dados históricos, detetando padrões e tendências e prevendo eventos futuros. O objetivo final é apoiar a tomada de decisão, a curto e médio prazo, para mitigar o fenómeno. Para tal, a plataforma considera a capacidade de carga dos locais alvo no cálculo da densidade de apinhamento. A integração de dados de diversas fontes é conseguida com uma arquitetura extensível, à base de conetores. São descritos três cenários de utilização da plataforma, relativos a eventos anuais de grande apinhamento. Dois deles, no município de Lisboa, baseados em dados de uma rede móvel disponibilizados pela iniciativa *LxDataLab*. O terceiro, em Melbourne na Austrália, utilizando dados públicos de uma rede de sensores de movimento designada de *Pedestrian Counting System*. É ainda descrita uma experiência de avaliação da usabilidade da plataforma proposta, usando o NASA-TLX.

**Palavras-chave:** turismo; visualização geo-temporal; apinhamento turístico; NASA-TLX

**Introdução**

Seja qual for a capacidade de carga dos pontos de interesse, os respetivos gestores turísticos devem planear, monitorizar e analisar os dados da sua visitação para prevenir ou mitigar os efeitos bem conhecidos do apinhamento ou sobrelotação turística, como a degradação da experiência dos visitantes, impactando negativamente o ambiente e a população local, o que pode a prazo tornar insustentável o turismo em destinos populares (Peeters et al. 2021).

Para analisar retrospetivamente, monitorizar e planear ações de mitigação deste fenómeno, estamos a desenvolver, no âmbito do projeto europeu RESETTING, uma plataforma de visualização geo-temporal, para explorar dados históricos do apinhamento, visualizar padrões, compreender tendências e prever situações futuras. Refira-se que uma plataforma de visualização é adequada para auxiliar na análise dos dados, enquanto uma solução computacional totalmente automatizada seria limitante e não apoiaria adequadamente a análise exploratória (Munzner, 2014).

A plataforma desenvolvida permite visualizar dados históricos ou quase em tempo real, auxiliando na tomada de decisões a curto e médio prazo. Por exemplo, no caso das autoridades municipais, tais decisões podem incluir a mobilização de recursos de limpeza urbana, policiamento e emergência médica em eventos com previsão de apinhamento. No caso dos turistas, podem traduzir-se na recomendação de percursos e locais de visitação alternativos, igualmente interessantes, mas não tão apinhados, ou dos períodos temporais em que os locais mais populares devam ser evitados. Em suma, este tipo de plataforma é uma componente essencial para o desenvolvimento de cidades inteligentes (Li et al., 2019; Bibri & Krogstie, 2020), com especial ênfase nos seus visitantes.

Na próxima seção descrevem-se as funcionalidades básicas da plataforma. De seguida, é apresentada a arquitetura computacional da plataforma e três casos de estudo da sua utilização, bem como os resultados da avaliação de usabilidade e escalabilidade da plataforma. A terminar apresentam-se algumas conclusões e planos de trabalho futuro.

**Funcionalidades**

O *frontend* da plataforma contém três componentes interligados, visíveis nas Figuras 1, 2 e 3: linha do tempo deslizante (em cima), mapa cartográfico 3D com colunas (todo o fundo) e gráfico de evolução temporal (em baixo, à esquerda). Os dados visualizados representam a(s) métrica(s) do apinhamento selecionada(s) e estão referenciados geográfica e temporalmente. Quando os dados de apinhamento são gerados em tempo real, o *frontend* da plataforma recolhe novos dados assim que estes estão disponíveis na fonte de dados, utilizando para o efeito um mecanismo de publicação-subscrição.

A principal visualização é baseada num **mapa cartográfico tridimensional**, operando sobre dados abertos (*[OpenStreetMap](#)*), que se pode arrastar e girar, bem como variar a altura de observação e mudança de escala (*zoom*). Nos locais de captura de dados são sobrepostos cilindros, cuja altura e cor são inferidas através do mapeamento de métricas disponíveis, a selecionar pelo utilizador. Por exemplo, a altura pode corresponder ao número de dispositivos detetados em *roaming* (turistas) e a cor representar a densidade da mesma métrica, usando a capacidade de carga da zona correspondente como denominador. Esta capacidade de carga é calculada tal como em (Brito e Abreu & Sampaio de Almeida, 2021), com o algoritmo descrito em (Sampaio de Almeida, 2021).

O mapa referido mostra dados espaciais para um único instante no tempo. No entanto, ao variar esse instante na **linha de tempo deslizante**, é criada uma visualização cinemática. O fundo gráfico deste controlo deslizante é um gradiente de cores que codifica a soma dos valores para todo o mapa, ou apenas para a região de interesse selecionada pelo utilizador, como no caso dos polígonos retangulares de 200m de lado na cidade de Lisboa (Figura 1). A atribuição de tonalidade é relativa aos locais considerados, de acordo com uma escala de cores; o instante que tem a soma de valores mais baixa na linha temporal é atribuído a uma tonalidade 120 no [modelo HSL](#) (verde puro), enquanto aquele com a soma mais alta é atribuído a uma tonalidade 0 (vermelho puro). Ao passar com o cursor por cima do botão da linha de tempo deslizante, é mostrada uma etiqueta com a data e hora do momento que está a ser visualizado no mapa. Finalmente, no **gráfico de evolução temporal**, que também permite arrasto e mudança de escala, as linhas representam o valor integral instantâneo de cada uma das métricas selecionadas para todo o mapa ou, se for o caso, para a região de interesse selecionada. No caso de serem representadas duas métricas, este gráfico apresentará dois eixos das ordenadas (à esquerda e à direita do gráfico) cada um com a escala correspondente. O eixo das abcissas não é representado para tornar a visualização menos sobrecarregada, mas a data e hora de cada ponto é também observável numa etiqueta ao sobrepor-lhe o cursor. Uma demonstração detalhada das funcionalidades desta plataforma está disponível no [canal YouTube do RESETTING](#).

**Arquitetura computacional**

Tecnicamente, a plataforma é uma aplicação web. O *backend*, construído com a tecnologia *Python/[Flask](#)*, é composto por uma série de módulos a que chamamos conetores. Cada conector é responsável pela implementação de um dos três tipos de *endpoints*, que podem servir para carregar dados históricos, carregar dados em tempo real ou para obter uma lista de locais especificados em formato *[GeoJSON](#)*. Os *endpoints* são instanciados de acordo com um ficheiro de configuração que contém informações tais como credenciais de base de dados e esquema associado. O mesmo ficheiro permite ainda especificar expressões de métricas derivadas com base em métricas preexistentes.

Esta arquitetura flexível permite conectar a plataforma a novas fontes de dados sem modificar o código do *frontend*. Foram implementados conetores para diferentes fontes de dados abertas: *InfluxDB* (base de dados de séries temporais), *Apache Kafka* (plataforma de transmissão de eventos) e *Opendatasoft* (plataforma de integração de dados). Com base nesses conetores, construíram-se três integrações com diferentes fontes de dados: os de uma operadora de rede móvel, os de uma rede de sensores de movimento e os de sensores Wi-Fi desenvolvidos também no âmbito do projeto *RESETTING*, este último caso descrito em (Mestre Santos, 2024).

O *frontend* foi construído com base nas tecnologias *Javascript*/*React*. Tal como para o *backend*, o *frontend* precisa ser configurado para funcionar. Dentro das configurações necessárias, a configuração de métricas disponíveis é particularmente relevante. Para cada métrica, existe uma propriedade denominada "cap" que representa o que seria considerado um valor muito alto para essa métrica. Tal é necessário para dimensionar corretamente os cilindros na visualização principal aos valores da métrica. Quando há uma métrica mapeada à cor dos cilindros, a propriedade "cap" é usada na determinação da cor.

Para que a plataforma possa estar hospedada na nuvem, com acesso público na internet, mas mantendo a confidencialidade dos dados, foi implementado um mecanismo de controlo que apenas permite o acesso a utilizadores autorizados, com chaves encriptadas guardadas numa base de dados.

O código-fonte da plataforma está disponível como código aberto no repositório *GitHub*.

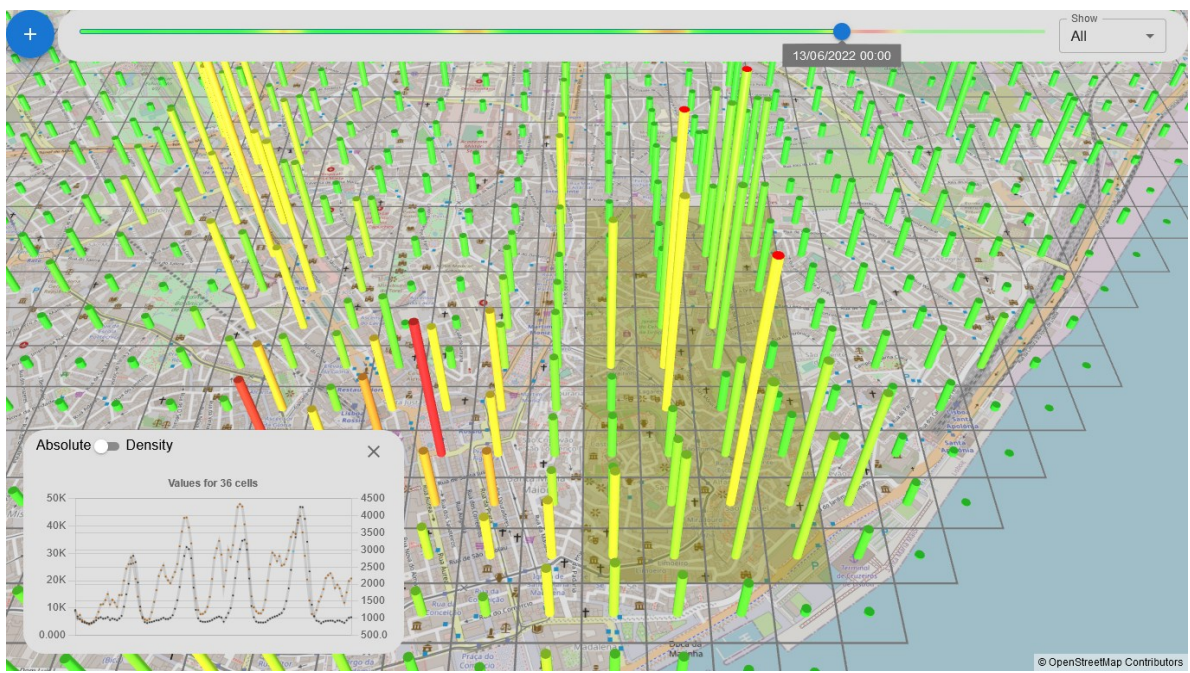

Figura 1: Identificação do apinhamento durante a comemoração dos Santos Populares em Lisboa.

**Casos de estudo**

Para exemplificar a utilização da plataforma, analisam-se três eventos anuais com alto apinhamento: dois em Lisboa (Santos Populares e o *Rock in Rio Lisboa*) e outro em Melbourne (passagem do ano).

No caso de Lisboa são usados dados da rede móvel da Vodafone, disponibilizados pelo Laboratório de Dados Urbanos de Lisboa (LxDataLab), incluindo métricas como o total de dispositivos detetados, ou só os que estejam em *roaming* (tipicamente dos turistas), em cada quadrícula de 200x200m, a cada 5 minutos. Os dados são provenientes da triangulação da contagem de dispositivos detetados por cada antena da rede celular.

Em relação aos Santos Populares de 2022, em Lisboa, poderíamos esperar que o número máximo de dispositivos detetados em Lisboa ocorresse durante o fim de semana alargado, entre 10 e 13 de junho. No entanto, podemos verificar através do gráfico de linhas que o pico ocorre no dia 9, indiciando que muitas pessoas regressam às suas cidades-dormitório ou viajam durante o fim de semana alargado. Ao selecionar a opção de mapear a cor dos cilindros à densidade de dispositivos tendo em conta a capacidade de carga, é possível ter uma perceção mais imediata dos focos de aglomeração. Para o pico no dia 9, verificam-se poucos cilindros com cor mais intensa, sendo que se destacam zonas de escritórios e o aeroporto. Passando à análise do evento propriamente dito, verificamos no gráfico de evolução temporal um máximo local na noite de 12 para 13 de junho. Ao arrastar o controlo deslizante para momentos desse intervalo temporal, verificamos que as zonas com mais apinhamento são a Avenida da Liberdade (marchas populares) e Alfama (arraiais), como seria expectável. Se quisermos saber qual o momento em que essas zonas específicas estiveram mais apinhadas, podemos selecioná-las desenhando polígonos no mapa. Fazendo isso, podemos verificar no gráfico que o pico nessa zona ocorreu próximo da meia-noite (Figura 1).

O *Rock in Rio Lisboa 2022* ocorreu no Parque da Bela Vista, também em Lisboa, durante dois fins de semana (18, 19, 25 e 26 de junho). Após carregar os dados correspondentes ao intervalo de 18 a 27 de junho, selecionamos o parque da Bela Vista, o que permite verificar no gráfico de linhas quatro curvas semelhantes, um para cada dia do evento, sendo que as curvas do segundo fim de semana têm picos maiores, chegando ao máximo global no dia 26 por volta das 23:30 (Figura 2). O conjunto de dados usado possui uma métrica que corresponde à duração média de permanência dos dispositivos numa quadrícula. O valor médio dessa métrica para a área selecionada durante o pico de aglomeração é de 9 minutos. Isto significa que houve muito movimento entre quadrículas, enquanto poderíamos esperar o contrário durante a realização de concertos. Podemos verificar ainda, usando o gráfico de linhas, que este valor não se trata de uma anomalia, sendo que essa métrica permanece com valores baixos durante o período do festival.

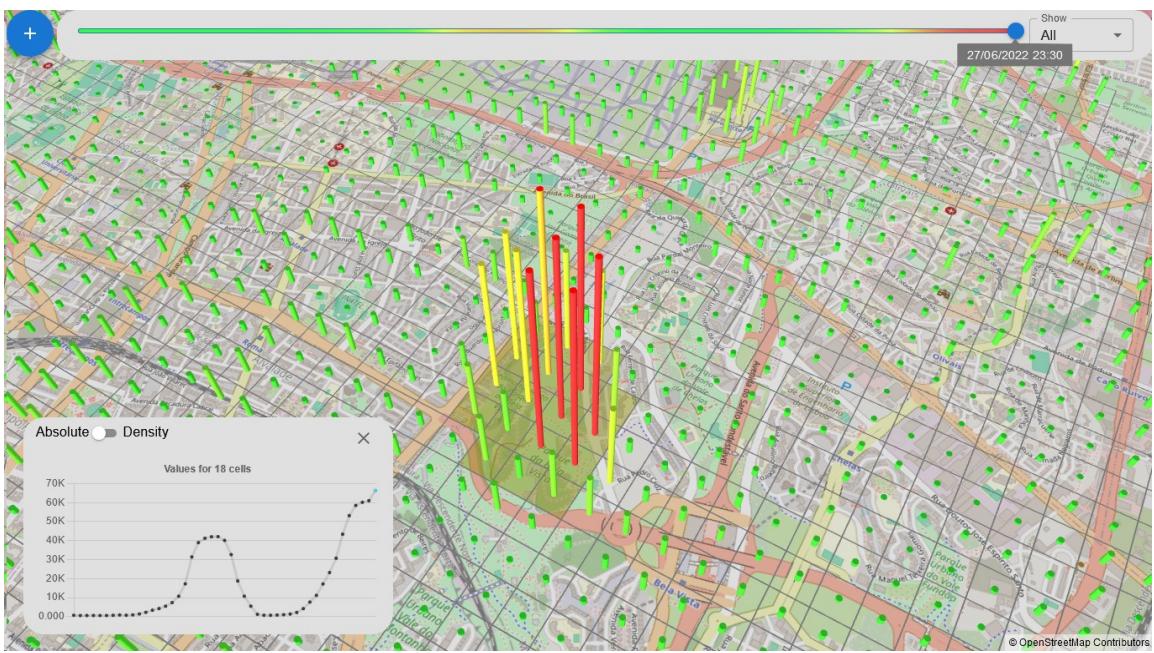

Figura 2: Identificação do apinhamento durante o *Rock in Rio Lisboa* 2022.

No caso de Melbourne são usados dados do *Pedestrian Counting System*, o qual inclui atualmente sensores de movimento em 93 locais da cidade, selecionados com base em três critérios – atividade comercial e de eventos, uso regular pelos pedestres e fluxo de saída e entrada nessas áreas. Os sensores estão instalados em pontos elevados, como no topo de postes de iluminação, para cobrir zonas de contagem na calçada abaixo. As contagens de pedestres são armazenadas localmente e transferidas para um servidor central a cada 10 minutos através de uma caixa de comunicação 3G.

Analisámos a passagem de ano de 2022 para 2023, um dos eventos com maior apinhamento que ocorrem na zona abrangida pelos sensores. Usando a plataforma para visualizar estes dados, verificamos a existência de um pico de apinhamento no dia 31 de dezembro às 22 horas, com cerca de 75000 pedestres detetados (Figura 3). O pico do dia seguinte, por sua vez, não chegou a 30000.

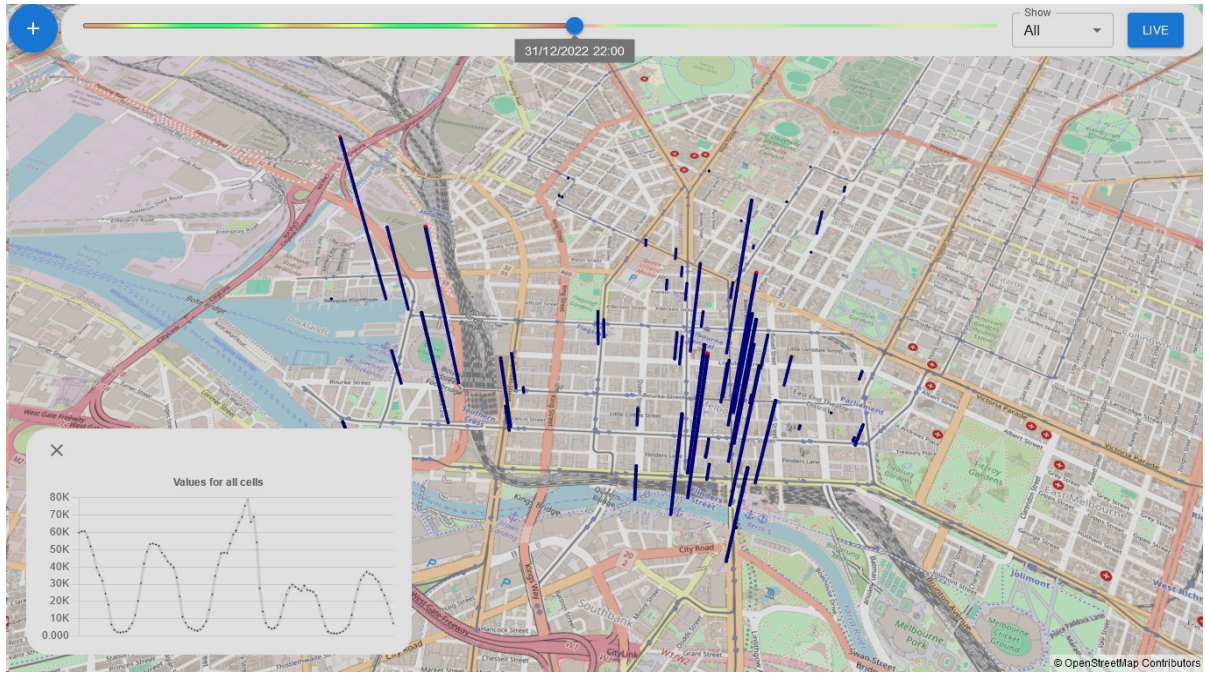

Figura 3: Utilização da plataforma com dados da cidade de Melbourne.

**Avaliação de usabilidade e escalabilidade**

Para avaliar a usabilidade da plataforma, elaborou-se uma experiência padronizada individual de três passos:

i. demonstração ao participante das várias funcionalidades da plataforma;
ii. realização pelo participante de uma tarefa na plataforma para responder a uma pergunta sobre o apinhamento no *Rock in Rio Lisboa 2022*;
iii. aplicação ao participante de um questionário amplamente difundido (Hart, 2006), o *NASA-Task Load Index* (NASA-TLX), para medir a carga mental de trabalho que o sujeito atribui à realização da tarefa referida.

Participaram nesta experiência 34 voluntários, com diferentes níveis de literacia tecnológica. A média do valor obtido para o NASA-TLX, numa escala de 0 a 100, foi de **33.45**. De acordo com a análise apresentada em (Grier, 2015) este valor fica pouco acima do quartil 25 para tarefas feitas num computador. Por outras palavras, a utilização da plataforma foi considerada de baixa a média dificuldade.

Fizemos também uma análise ao desempenho da plataforma. A visualização dos dados no *frontend* é quase instantânea, sendo que o constrangimento no processamento se dá ao nível da recolha e processamento dos dados no *backend*, onde o tempo de processamento aumenta de forma não linear com a quantidade de dados envolvida.

## Conclusões

A plataforma proposta permite a análise exploratória de dados geo-temporais de apinhamento, tanto históricos como em tempo real. Neste artigo foram apresentados vários cenários da sua utilização e apresentados resultados da avaliação da sua usabilidade e escalabilidade.

O *frontend* da plataforma é uma aplicação web que proporciona uma experiência de utilização com uma carga mental relativamente baixa, avaliada experimentalmente com o NASA-TLX por 34 voluntários. O seu *backend* possui uma arquitetura extensível baseada em conectores, permitindo a recolha de dados de novos tipos de fontes através da criação de novos módulos, sem ser necessário alterar o código do *frontend*.

Esta plataforma de visualização geo-temporal está em fase de desenvolvimento. Como trabalho futuro, para além da componente de previsão por implementar, contamos também demonstrar com *stakeholders* concretos como é que a plataforma pode ajudar a compreender, prever e planear ações de mitigação do apinhamento turístico.

## Agradecimentos

## Referências: